\documentclass[preprint,showpacs,preprintnumbers,amsmath,amssymb]{revtex4}

\usepackage{graphicx}
\usepackage{epsfig}		
\usepackage{dcolumn}
\usepackage{bm}

\def\0{\mbox{\tiny $0$}}
\def\1{\mbox{\tiny $1$}}
\def\2{\mbox{\tiny $2$}}
\def\3{\mbox{\tiny $3$}}
\def\4{\mbox{\tiny $4$}}
\def\5{\mbox{\tiny $5$}}
\def\6{\mbox{\tiny $6$}} 
\def\7{\mbox{\tiny $7$}}
\def\8{\mbox{\tiny $8$}}
\def\9{\mbox{\tiny $9$}}
\def\R{\mbox{\tiny $R$}}
\def\T{\mbox{\tiny $T$}}

\def\bb{\mbox{\tiny $B$}}

\def\L{\mbox{\tiny $L$}}

\def\pl{\mbox{\tiny $+$}}

\begin{document}

\title{LIMITATIONS ON THE PRINCIPLE OF STATIONARY PHASE WHEN IT IS APPLIED TO TUNNELING ANALYSIS}

\author{A. E. Bernardini}
\email{alexeb@ifi.unicamp.br}
\affiliation{Instituto de F\'{\i}sica Gleb Wataghin, UNICAMP,\\
PO Box 6165, 13083-970, Campinas, SP, Brasil.}

\date{\today}

\begin{abstract}
Using a recently developed procedure - multiple wave packet decomposition - here we study the phase time formulation for tunneling/reflecting particles colliding with a potential barrier.
To partially overcome the analytical difficulties which frequently arise when the stationary phase method is employed for deriving phase (tunneling) time expressions, we present a theoretical exercise involving a symmetrical collision between two identical wave packets and an one-dimensional rectangular potential barrier.
Summing the amplitudes of the reflected and transmitted waves - using a method we call multiple peak decomposition - is shown to allow reconstruction of the scattered wave packets in a way which allows the stationary phase principle to be recovered.
\end{abstract}

\pacs{03.65.Xp}
\keywords{Stationary Phase Method - Phase Times - Tunnel Effect}
\date{\today}
\maketitle

Recently, a series of experimental results \cite{Nim92,Ste93,Chi98,Hay01}, some of them confirming the possibility of superluminal tunneling speeds for photons, have revived an interest in the tunneling time analysis \cite{Olk04,Pri03,Lan94,Olk92,Hau89}.
On the theoretical front, people have tried to introduce quantities that have the dimension of time and can somehow be associated with the passage of the particle through the barrier or, strictly speaking, with the definition of the tunneling time.
These proposals have led to the introduction of several {\em time} definitions \cite{Olk04,Baz67,But83,Hau87,Fer90,Yuc92,Hag92,Bro94,Sok94,Olk95,Jak98,Olk02}, some of which can be organized into three groups.
(1) The first group comprises a time-dependent description in terms of wave packets where some features of an incident packet and the comparable features of the transmitted packet are 
utilized to describe a quantifiable {\em delay} as a tunneling time \cite{Hau89}.
(2) In the second group the tunneling times are computed based on averages over a set of kinematical paths, whose distribution is supposed to describe the particle motion inside a barrier.
In this case, Feynman paths are used like real paths to calculate an average tunneling time with the weighting function $\exp{[i\, S\, x(t)/\hbar]}$, where $S$ is the action associated with the path $x(t)$ (where $x(t)$ represents the Feynman paths initiated from a point on the left of the barrier and ending at another point on the right of it \cite{Sok87}).
The Wigner distribution paths \cite{Bro94}, and the Bohm approach \cite{Ima97,Abo00} are included in this group.
(3) In the third group we notice the introduction of a new degree of freedom, constituting a physical clock for the measurements of tunneling times.
This group comprises the methods with a Larmor clock \cite{But83} or an oscillating barrier \cite{But82}.
Separately, standing on itself is the {\em dwell} time defined by the interval during which the incident flux has to exist and act, to provide the expected accumulated particle storage, inside the barrier \cite{Lan94}.

There is no general agreement \cite{Olk04,Olk92} among the above definitions about the meaning of tunneling times (some of the proposed tunneling times are actually traversal times, while others seem to represent in reality only the spread of their distributions) and about which, if any, of them is the proper tunneling time \cite{Olk04}.
In the context where we intend to work on, the tunneling mechanism is embedded by theoretical constructions involving analytically-continuous {\em gaussian}, or infinite-bandwidth step pulses to examine the tunneling process.
Nevertheless, such holomorphic functions do not have a well-defined front in a manner that the interpretation of the wave packet speed of propagation becomes ambiguous.
Moreover, infinite-bandwidth signals cannot propagate through any real physical medium (whose transfer function is therefore finite) without pulse distortion, which also leads to ambiguities in determining the propagation velocity during the tunneling process.
For instance, some of the barrier traversal time definitions lead, under tunneling time conditions, to very short times, which can even become negative.
It can precipitately induces an interpretation of violation of simple concepts of causality.
Otherwise, negative speeds do not seem to create problems with causality, since they were predicted both within special relativity and within quantum mechanics \cite{Olk95}.
A possible explanation of the time advancements related to the negative speeds can come, in any case, from consideration of the very rapid spreading of the initial and transmitted wave packets for large momentum distribution widths.
Due to the similarities between tunneling (quantum) packets and evanescent (classical) waves, exactly the same phenomena are to be expected in the case of classical barriers\footnote{In particular, we could mention the analogy between the stationary Helmholtz equation for an electromagnetic wave packet - in a waveguide, for instance - in the presence of a {\em classical} barrier and the stationary Schroedinger equation, in the presence of a potential barrier \cite{Lan94,Jak98,Nim94}).}.
The existence of such negative times is predicted by relativity itself based on its ordinary postulates \cite{Olk04}, and they appear to have been experimentally detected in many works \cite{Gar70,Chu82}.

In this extensively explored scenario, the first group quoted above contains the so-called phase times \cite{Boh52,Wig55,Con70} which are obtained when the stationary phase method (SPM) \cite{PBE} is employed for obtaining the times related to the motion of the wave packet spatial centroid.
Generically speaking, the SPM essentially enables us to parameterize some subtleties of several quantum phenomena, such as tunneling \cite{Hau89,Ste93,Bro94}, resonances \cite{Con68,Bra70,Sok94B}, incidence-reflection and incidence-transmission interferences \cite{Per01} as well as the Hartman effect \cite{Har62} and its {\em superluminal} traversal time interpretation \cite{Olk04,Lan94,Jak98}.
In fact, it is the simplest and most usual approximation method for describing the group velocity of a wave packet in a quantum scattering process represented by the collision of a particle with a potential barrier \cite{Olk04,Lan94,Hau87,Wig55,Har62,Ber04}.

In the following study we will concentrate on some incompatibilities that appear when the SPM is utilized for deriving tunneling times.
After quantifying the restrictive conditions for the use of the method, at the end of our analysis, we discuss a theoretical exercise involving a symmetrical collision between two identical wave packets and an one-dimensional rectangular potential barrier.
We demonstrate that by summing the amplitudes of the reflected and transmitted waves in the scope of what we denominate a multiple peak decomposition analysis \cite{Ber04}, we can recompose the scattered wave packets in a way that the analytical conditions for the SPM applicability are totally recovered.

The SPM can be successfully applied for describing the movement of the center of a wave packet constructed in terms of a momentum distribution $g(k - k_{\0})$ which has a pronounced peak around $k_{\0}$. 
By assuming that the phase that characterizes the propagation varies smoothly around the maximum of $g(k - k_{\0})$, the stationary phase condition enables us to calculate the position of the peak of the wave packet (highest probability region to find the propagating particle).
With regard to the tunneling phenomenon, the method is usually applied to find the position of a wave packet that traverses a potential barrier.
For the case in which we consider a rectangular potential barrier $V(x)$, $V(x) = V_{\0}$ if $x \in \mbox{$\left[- L/2, \, L/2\right]$}$
and $V(x) = 0$ if $x \in\hspace{-0.3cm}\slash\hspace{0.1cm}\mbox{$\left[- L/2, \, L/2\right]$}$,
\small\begin{equation}
V(x) = \left\{\begin{array}{cll} V_o && ~~~~x \in \mbox{$\left[- L/2, \, L/2\right]$}\\ &&\\ 0&& ~~~~x \in\hspace{-0.3cm}\slash\hspace{0.1cm}\mbox{$\left[- L/2, \, L/2\right]$}\end{array}\right.
\label{2p60}
\end{equation}\normalsize
it is well known that the transmitted wave packet solution ($x \geq L/2 $) calculated by means of the Schroedinger formalism is 
given by \cite{Coh77}
\small\begin{equation}
\psi^{\T}(x,t) = \int_{_{\0}}^{^{w}}\frac{dk}{2\pi} \, g(k - k_{\0}) \, |T(k, L)|\,
\exp{\left[ i \, k \,(x - L/2) - i \, \frac{k^2}{2\,m} \, t +
 i \,\Theta(k, L)\right]}.
\label{1}
\end{equation}\normalsize
In case of tunneling, the transmitted amplitude and the phase shift are respectively given by
\small\begin{equation}
|T(k, L)| =
\left\{1+ \frac{w^4}{4 \, k^2 \, \rho^{\2}(k)}
\sinh^2{\left[\rho(k)\, L \right]}\right\}^{-\frac{1}{2}},
\label{501}
\end{equation}\normalsize
and
\small\begin{equation}
\Theta(k, L) = \arctan{\left\{\frac{2\, k^2 - w^2}
{k \, \rho(k)}
\tanh{\left[\rho(k) \, L \right]}\right\}}, 
\label{502}
\end{equation}\normalsize
for which we have made explicit the dependence on the barrier length $L$, and we have adopted $\rho(k) = \left(w^2 - k^2\right)^{\frac{1}{2}}$ with $w = \left(2\, m \,V_{\0}\right)^{\frac{1}{2}}$ and $\hbar = 1$.
By not considering any eventual distortion that $|T(k, L)|$ could cause to the supposedly symmetric function $g(k - k_{\0})$, the stationary phase condition is indiscriminately applied to the phase (\ref{1}) leading to
\begin{eqnarray}
\frac{d}{dk}\left.\left\{k \,(x - L/2) - \frac{k^2}{2\,m} \, t 
+ \Theta(k, L)\right\}\right|_{_{k = k_{\mbox{\tiny max}}}} 
&=&0\nonumber\\
 ~~~~\Rightarrow
x - L/2 - \frac{k_{\mbox{\tiny max}}}{m} \, t + 
\left.\frac{d\Theta(k, L)}{dk}\right|_{_{k = k_{\mbox{\tiny max}}}} = 0.
\label{3}
\end{eqnarray}
The above result is frequently adopted for calculating the transit time $t_{T}$ of a transmitted wave packet when its peak emerges at $x = L/2$,
\small\begin{equation}
t_{T} =\frac{m}{k_{\mbox{\tiny max}}}\left.\frac{d\Theta(k, \alpha_{(\L)})}{dk}\right|_{_{k = k_{\mbox{\tiny max}}}} =
\frac{2\,m \, L}{k_{\mbox{\tiny max}} \,\alpha }
\left\{\frac{w^4\,\sinh{(\alpha)}\cosh{(\alpha)}
-\left(2\, k_{\mbox{\tiny max}}^2 - w^2 \right)k_{\mbox{\tiny max}}^2 \,\alpha }
{4\, k_{\mbox{\tiny max}}^2 \,\left(w^2 - k_{\mbox{\tiny max}}^2 \right)  +
w^4\,\sinh^2{(\alpha)}}\right\}
\label{4}
\end{equation}\normalsize
where we have defined the parameter $\alpha = \left(w^2 - k_{\mbox{\tiny max}}^2 \right)^{\frac{1}{2}}\, L$.
The concept of {\em opaque} limit is introduced when we assume that $k_{\mbox{\tiny max}}$ is independent of $L$ and then we make $\alpha$ tend to $\infty$ \cite{Jak98}.
In this case, the transit time can be rewritten as 
\small\begin{equation}
t^{^{OL}}_T = \frac{2\,m}{k_{\mbox{\tiny max}}\,\rho(k_{\mbox{\tiny max}})}.
\label{5}
\end{equation}\normalsize
In the literature, the value of $k_{\mbox{\tiny max}}$ is frequently approximated by $k_{\0}$, the maximum of $g(k - k_{\0})$, which, in fact, does not depend on $L$ and could lead us to the {\em superluminal} transmission time interpretation \cite{Jak98,Olk92,Esp03}.
To clear up this point, we notice that when we take the so called {\em opaque} limit in Eq.~(\ref{5}), with $L$ going to $\infty$ and
$w$ fixed as well as with $w$ going to $\infty$ and $L$ fixed, with $k_{\0} < w$ in
both cases, the expression (\ref{5}) leads to times corresponding
to a transmission process performed with velocities larger than $c$ \cite{Jak98}.

Such a {\em superluminal} interpretation was extended to the study of quantum tunneling through two successive barriers separated by a free region \cite{Olk02}.
In this approach, the total traversal time should be independent of the barrier widths and of the distance between the barriers.
In a subsequent analysis, the same technique was applied to a problem with multiple successive barriers where the tunneling process was designated as a highly non-local phenomenon \cite{Esp03}.

It would be perfectly acceptable to consider $k_{\mbox{\tiny max}} = k_{\0}$ for the application of the stationary phase condition if the momentum distribution $g(k - k_{\0})$ centered at $k_{\0}$ was not modified by any boundary condition.
That is the case of the incident wave packet before the collision with the potential barrier.
In this sense, and in the context of the above quoted theoretical results, our criticism is concerned with the way of obtaining all the above results for the transmitted wave packet.
It has not taken into account the bounds and enhancements imposed by the analytical form of the transmission coefficient.

To perform the correct analysis, we should calculate the correct value of $k_{\mbox{\tiny max}}$ to be substituted in Eq.~(\ref{4}) before taking the {\em opaque} limit.
We are thus obliged to consider the relevant amplitude for the transmitted wave as the product of a symmetric momentum distribution $g(k - k_{\0})$, which describes the {\em incoming} wave packet, by the modulus of the transmission amplitude $T(k, L)$, which is a crescent function of $k$.
The maximum of this product representing the transmission modulating function would be given by the solution of the equation
\begin{eqnarray}
g(k - k_{\0}) \,\left|T(k, L)\right|\,\left[\frac{g^{\prime}(k - k_{\0})}{g(k - k_{\0})}+
\frac{\left|T(k, L)\right|^{\prime}}{\left|T(k, L)\right|}\right] &=& 0.
\label{3p42B}
\end{eqnarray}
Obviously, the peak of the modified momentum distribution is shifted to the right of $k_{\0}$ so that $k_{\mbox{\tiny max}}$ has to be found in the interval $]k_{\0}, w[$.
Moreover, we can demonstrate by the numerical results of {\it Table 1} that $k_{\mbox{\tiny max}}$ presents an implicit dependence on $L$.
For obtaining the {\it Table 1} data we have found the maximum of $g(k - k_{\0}) \, |T(k, L)|$ by assuming a {\em gaussian} distribution $g(k - k_{\0}) = \left(\frac{a^2}{2 \, \pi}\right)^{\frac{1}{4}}\exp{\left[-\frac{a^2 (k -k_{\0})^2}{4}\right]}$ almost completely comprised in the interval $[0, w]$.

By increasing the value of $L$ with respect to the wave packet width $a$, the value of $k_{\mbox{\tiny max}}$ obtained from the numerical calculations to be substituted in Eq.~(\ref{4}) also increases up to $L$ reaches certain values for which the modified momentum distribution becomes unavoidably distorted.
In this case, the relevant values for $k$ are concentrated in the neighborhood of the upper boundary value $w$.
We shall show in the following that the value of $L$, which sets up the distortion the momentum distribution can be analytically obtained in terms of $a$.
\begin{table}[ht]
\begin{minipage}{14cm}
\begin{center}
\caption{The values of $k$ numerically obtained in correspondence with the increasing of the barrier extension $L$.
The values are calculated in terms of the wave packet width $a$ for different values of the potential barrier height expressed in terms of $w \, a$.
We have fixed the incoming momentum by setting $k_{\0} \, a = 1$.}
\begin{tabular}{c|ccccccc}
\hline
\hline
~~~~~~~~~$w \, a$& 1.5 & 2.0 & 4.0 & 6.0 & 8.0 & 10 & 20 \\
$ L / a $ &&&&&&&\\
\hline\hline
0.00 & 1.0000& 1.0000& 1.0000& 1.0000& 1.0000& 1.0000& 1.0000\\
0.10 & 1.0235& 1.0648& 1.3799& 1.6769& 1.8547& 1.9397& 2.0051\\
0.20 & 1.0794& 1.1825& 1.6571& 1.9178& 2.0000& 2.0204& 2.0203\\
0.30 & 1.1478& 1.3001& 1.8430& 2.0289& 2.0562& 2.0551& 2.0342\\
0.40 & 1.2196& 1.4116& 1.9874& 2.1025& 2.0986& 2.0857& 2.0484\\
0.50 & 1.2921& 1.5194& 2.1155& 2.1668& 2.1399& 2.1170& 2.0628\\
0.60 & 1.3649& 1.6266& 2.2429& 2.2314& 2.1828& 2.1495& 2.0775\\
0.70 & 1.4383& 1.7360& 2.3819& 2.3002& 2.2281& 2.1834& 2.0925\\
0.80 &    $*\footnote{For the values of $L$ marked with $*$, we can demonstrate by means of Eqs.~(\ref{b}-\ref{c}) that the modulated momentum distribution has already been completely distorted. In this case, the maximum has no meaning in the context of the applicability of the method of stationary phase.}
            $& 1.8489& 2.5466& 2.3751& 2.2761& 2.2188& 2.1078\\
0.90 &    $*$& 1.9646& 2.7627& 2.4578& 2.3272& 2.2558& 2.1234\\
1.00 &    $*$&    $*$& 3.1137& 2.5504& 2.3818& 2.2947& 2.1392\\
\hline\hline
\end{tabular} 
\end{center}
\end{minipage}
\end{table}\normalsize

Now, if we take the {\em opaque} limit of $\alpha$ by fixing $L$ and increasing $w$, the above results immediately ruin the {\em superluminal} interpretation upon the result of Eq.~(\ref{4}), since $t^{^{OL}}_T$ tends to $\infty$ when $k_{\mbox{\tiny max}}$ is substituted by $w$.
Otherwise, when $w$ is fixed and $L$ tends to $\infty$, the parameter $\alpha$ calculated at $k = w$ becomes indeterminate.
The transit time $t_T$ still tends to $\infty$ but now it exhibits a peculiar dependence on $L$, which can be easily observed by defining the auxiliary function
\small\begin{equation}
G(\alpha) =  
\frac{\sinh{(\alpha)}\cosh{(\alpha)} - \alpha}
{\sinh^2{(\alpha)}}.
\label{11}
\end{equation}\normalsize
When $\alpha \gg 1$,
the transmission time assumes infinite values
\small\begin{equation}
t^{\alpha}_T = \frac{2\, m\, L}{w \, \alpha}\,
G(\alpha)~~ \Rightarrow~~ t^{\alpha}_T \approx  \frac{2\, m }{w \, \left(w^2 - k^2 \right)^{\frac{1}{2}}} \rightarrow \infty.
\label{13}
\end{equation}\normalsize
with an asymptotic dependence on $\left(w^2 - k^2 \right)^{-\frac{1}{2}}$.
Only when $\alpha$ tends to $0$ we have an explicit linear dependence on $L$ given by
\small\begin{equation}
t^{0}_T = \frac{2\, m \, L}{w}\,\lim_{\alpha \rightarrow 0}
{\left\{\frac{G(\alpha)}{\alpha}\right\}}
= \frac{4\, m \, L}{3 \, w}
\label{14}
\end{equation}\normalsize

In addition to the above results, the transmitted wave must be carefully studied in terms of the ratio between the barrier extension $L$ and the wave packet width $a$.
For very thin barriers, i. e. when $L$ is much smaller than $a$, the modified transmitted wave packet presents substantially the same form of the incident one.
For thicker barriers, but yet with $L < a$, the peak of the {\em gaussian} wave packet modulated by the transmission coefficient is shifted to higher energy values, i. e.  $k_{\mbox{\tiny max}} > k_{\0}$ increases with $L$.
For very thick barriers, i. e. when $L > a$, we are able to observe that the form of the transmitted wave packet is badly distorted with the greatest contribution coming from the Fourier components corresponding to the energy $w$ just above the top of the barrier in a kind of
{\em filter effect}.
We observe that the quoted distortion starts to appear when the modulated momentum distribution presents a {\em local maximal} point at $k = w$ which occurs when
$\left.\frac{d}{dk}\left[g(k - k_{\0}) \,
\left|T(k, L)\right|\right]\right|_{k = w} > 0.$
Since the derivative of the {\em gaussian} function $g(k - k_{\0})$ is negative at $k = w$, the previous relation gives
\small\begin{equation}
- \frac{g^{\prime}(w - k_{\0})}{g(w - k_{\0})} < \lim_{k \rightarrow w}
{\left[\frac{T^{\prime}(k, L)}{T(k, L)}\right]}
= \frac{w \, L^2}{4}\frac{\left(1 + \frac{w \, L^2}{3}\right)}
{\left(1 + \frac{w \, L^2}{4}\right)} < \frac{w \, L^2}{3}
\label{b}
\end{equation}\normalsize
which effectively represents the inequality
\small\begin{equation}
\frac{a^2}{2}\,(w - k_{\0}) < \frac{w \,L^2}{3}~~\Rightarrow~~
L > \sqrt{\frac{3}{2}}\,a\, \left(1 - \frac{k_{\0}}{w}\right).
\label{c}
\end{equation}\normalsize

\begin{figure}[h]
\centerline{\psfig{file=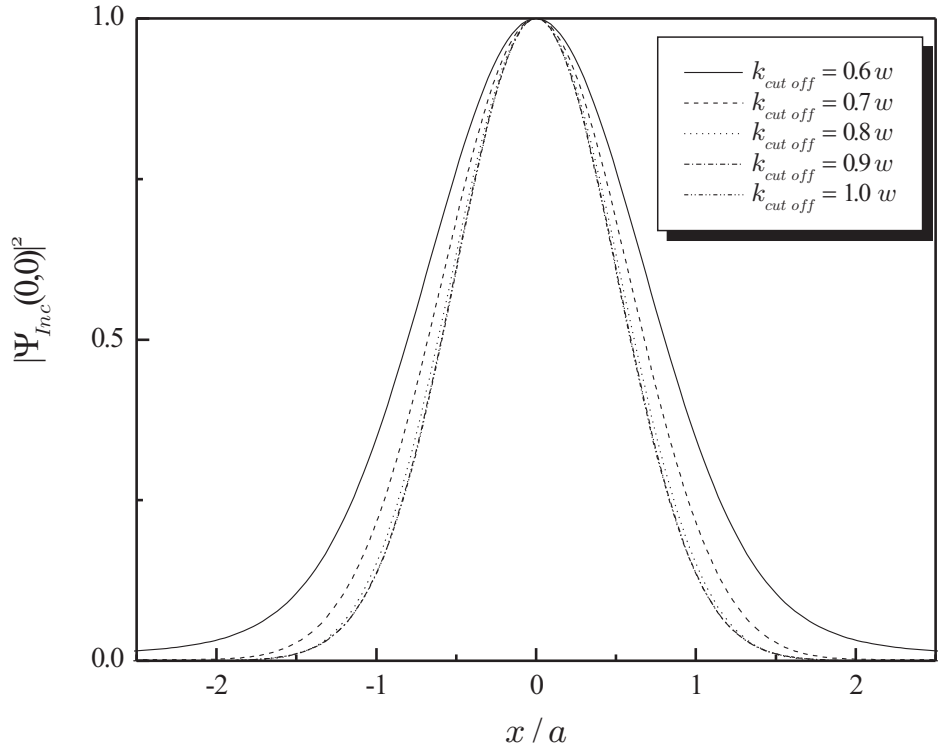,width=14cm}}
\caption{Dependence of the wave packet shape on the {\em cut off} value of a momentum distribution centered around $k_{o} = 0.5 w$ with the values of $k$ comprised between $0$ and $k_{\mbox{\tiny \em{cut off}}}$.\label{fig2}}
\end{figure}

Due to the {\em filter effect}, the amplitude of the transmitted wave is essentially composed by the plane wave components of the front tail of the {\em incoming} wave packet that reaches the first barrier interface before the peak arrival.
Meanwhile, only whether we had {\em cut} the momentum distribution {\em off} at a value of $k$ smaller than $w$, i. e. $k \approx (1 - \delta) w$, the {\em superluminal} interpretation of the transition time (\ref{5}) could be recovered.
In this case, independently of the way that $\alpha$ tends to $\infty$, the value assumed by the transit time would be approximated by $t^{\alpha}_{T} \approx 2 \,m / w \, \delta$, which is a finite quantity.
Such a finite value would confirm the hypothesis of {\em superluminality}.
However, the {\em cut off} at $k \approx (1 - \delta) w$ increases the amplitude of the tail of the incident wave as we can observe in Fig.~\ref{fig2}.
It means that the contribution of wave packet tail for the final composition of the transmitted wave is put on the same level with the contribution of the peak of the incident wave.
Consequently, an ambiguity in the definition of the {\em arrival} time is created.

To summarize, at this point we are particularly convinced that the use of a step-discontinuity to analyze signal transmissions in tunneling processes deserves a more careful analysis than the immediate application of the stationary phase method.
The point is that we cannot find an analytic-continuation between the {\em above-barrier} case solutions and the {\em below-barrier} case solutions.
By assuming the factual influence of the amplitude of the transmitted wave, we may introduce an alternative analysis where we consider the possibility of using the multiple peak decomposition technique developed for the above barrier diffusion problem \cite{Ber04}.
By means of such an experimentally verifiable exercise, we shall be able to understand how the {\em filter effect} can analytically affect the calculations of transit times in the tunnel process.

In the framework of the multiple peak decomposition \cite{Ber04}, we suggest a suitable way for comprehending the conservation of probabilities for a very particular scattering configuration where the asymmetric aspects above discussed can be totally eliminated.
In order to recover the scattered momentum distribution symmetry conditions for accurately applying the SPM, we assume a symmetrical colliding configuration of two wave packets traveling in opposite directions. 
By considering the same rectangular barrier $V(x)$, we solve the Schroedinger equation for a plane wave component of momentum $k$ for two identical wave packets symmetrically separated from the origin $x = 0$.
At time $t = - (m L) /(2 k_{\0})$ chosen for mathematical convenience, we assume that they perform a totally symmetric simultaneous collision with the potential barrier.
The wave packet reaching the left(right) side of the barrier is represented by
\small\begin{equation}
\psi^{\L(\R)}(x,t) = \int_{_{\0}}^{^{\pl\infty}}dk \, g(k - k_{\0})\phi^{\L(\R)}(k,x)\, \exp{[- i \, E\,t]}.
\end{equation}\normalsize
Here we have assumed that the limits of the above integral can be naturally extended from the interval $[0,w]$ to the interval $[0, \infty]$ as a first approximation.
Its range of validity can be controlled by the choice of the width $\Delta k$ of the momentum distribution $g(k-k_{\0})$ (with $k_{\0} > 0$) with $\Delta k$ enhanced by the barrier's height $(V_{\0})$.
By assuming that $\phi^{\L(\R)}(k,x)$ are Schroedinger equation solutions, at the time $t = - (m L) /(2 k_{\0})$, i. e. when the wave packet peaks simultaneously reach the barrier, we can write
\small\begin{equation}
\phi^{\L(\R)}(k,x)=\left\{
\begin{array}{l l l l}
\phi^{\L(\R)}_{\1}(k,x) &=& \exp{\left[ \pm i \,k \,x\right]} + R^{\L(\R)}_{\bb}(k,L)\exp{\left[ \mp i \,k \,x\right]}&~~~~x < - L/2\, (x > L/2),\nonumber\\
\phi^{\L(\R)}_{\2}(k,x) &=& \alpha^{\L(\R)}_{\bb}(k)\exp{\left[ \mp\rho  \,x\right]} + \beta^{\L(\R)}_{\bb}(k)\exp{\left[ \pm\rho  \,x\right]}&~~~~- L/2 < x < L/2,\nonumber\\
\phi^{\L(\R)}_{\3}(k,x) &=& T^{\L(\R)}_{\bb}(k,L)\exp{ \left[\pm i \,k \,x\right]}&~~~~x > L/2 \, (x < - L/2) .
\end{array}\right.
\label{510}
\end{equation}\normalsize
where the upper(lower) sign is related to the index $L$($R$).
By assuming the conditions for the continuity of $\phi^{\L,\R}$ and their derivatives at $x = - L/2$ and $x = L/2$, after some mathematical manipulations, we can easily obtain
\small\begin{equation}
R^{\L,\R}_{\bb}(k,L) = \exp{\left[ - i \,k \,L \right]} \left\{\frac{\exp{\left[ i \, \Theta(k,L)\right]} \left[1 - \exp{\left[ 2\,\rho(k) \,L\right]}\right]}{1 - \exp{\left[ 2\,\rho(k) \,L\right]}\exp{\left[ i\, \Theta(k,L)\right]}}\right\}
\label{511}
\end{equation}\normalsize
and
\small\begin{equation}
T^{\L,\R}_{\bb}(k,L) = \exp{\left[ - i \,k \,L \right]} \left\{\frac{\exp{\left[\rho(k) \,L\right]}\left[1- \exp{\left[ 2\, i\, \Theta(k,L)\right]}\right]}{1 - \exp{\left[ 2\,\rho(k) \,L\right]}\exp{\left[ i\, \Theta(k,L)\right]}}\right\},
\label{512}
\end{equation}\normalsize
where $\Theta(k,L)$ is given by the Eq.~(\ref{502}) and $R^{\L}_{\bb}(k,L)$ and $T^{\R}_{\bb}(k,L)$ as well as $R^{\R}_{\bb}(k,L)$ and $T^{\L}_{\bb}(k,L)$  are intersecting each other.
By analogy with the procedure of {\em summing amplitudes} that we have adopted in the multiple peak decomposition scattering \cite{Ber04},
such a pictorial configuration obliges us to sum the intersecting amplitude of probabilities before taking their squared modulus in order to obtain
\small\begin{equation}
R^{\L,\R}_{\bb}(k,L)+ T^{\R,\L}_{\bb}(k,L) = \exp{\left[ - i \,k \,L \right]} \left\{\frac{\exp{\left[\rho(k) \,L\right]}+ \exp{\left[ i\, \Theta(k,L)\right]}}{1 + \exp{\left[ \rho(k) \,L\right]}\exp{\left[ i\, \Theta(k,L)\right]}}\right\}
 = \exp{\left\{ - i [k \,L + \varphi(k,L)]\right\}}
\label{513}
\end{equation}\normalsize
with
\small\begin{equation}
\varphi(k,L) = \arctan{\left\{\frac{2\,k\,\rho(k) \, \sinh{[\rho(k)\,L]}}{w^{\2} + \left(k^{\2}-\rho^{\2}(k)\right)\cosh{[\rho(k)\,L]}}\right\}}.
\label{514}
\end{equation}\normalsize\normalsize
From Eq.~(\ref{513}), it is important to observe that, differently from the previous standard tunneling analysis, by adding the intersecting amplitudes at each side of the barrier, we keep the original momentum distribution undistorted since $|R^{\L,\R}_{\bb}(k,L)+ T^{\R,\L}_{\bb}(k,L)|$ is equal to one.
At this point we recover the most fundamental condition for the applicability of the SPM.
It allows us to accurately find the position of the peak of the reconstructed wave packet composed by reflected and transmitted superposing components.
The phase time interpretation can be, in this case, correctly quantified in terms of the analysis of the {\em new} phase $\varphi(k, L)$.
By applying the stationary phase condition to the recomposed wave packets, the maximal point of the scattered amplitudes $g(k - k_{\0})|R^{\L,\R}_{\bb}(k,L)+ T^{\R,\L}_{\bb}(k,L)|$ are accurately given by $k_{\mbox{\tiny max}} = k_{\0}$ so that the traversal/reflection time or, more generically, the scattering time, results in
\small\begin{equation}
t^{^{\varphi}}_{T} =\frac{m }{k_{\0}}\left.\frac{d\varphi(k, \alpha_{(\L)})}{dk}\right|_{_{k = k_{\0}}} =
\frac{2\,m\, L}{k_{\0}\,\alpha}
\frac{w^{\2}\sinh{(\alpha)} - \alpha\,k^{\2}_{\0}}{2\,k^{\2}_{\0} - w^{\2} + w^{\2}\cosh^{\2}{(\alpha)}}
\label{515}
\end{equation}\normalsize
with $\alpha$ previously defined. 
It can be said metaphorically that the identical particles represented by both incident wave packets spend a time of the order of $ t^{^{\varphi}}_{T}$ inside the barrier before retracing its steps or tunneling.
In fact, we cannot differentiate the tunneling from the reflecting waves for such a scattering configuration.
The point is that we have introduced the possibility of improving the efficiency of the SPM in calculating reflecting and tunneling phase times, by studying a process where the conditions for applying the method are totally recovered.
We have demonstrated that the transmitted and reflected interfering amplitudes results in a unimodular function which just modifies the {\em envelop} function $g(k - k_{\0})$ by an additional phase.
The previously pointed out incongruities which cause the distortion of the momentum distribution $g(k - k_{\0})$ are completely eliminated in this case.
At the same time, one could argue about the possibility of extending such a result to the tunneling process established in a standard way.
We should assume that in the region inside the potential barrier, the reflecting and transmitting amplitudes should be summed before we compute the phase changes.
Obviously, it would result in the same phase time expression as represented by (\ref{515}).
In this case, the assumption of there (not) existing interference between the momentum amplitudes of the reflected and transmitted waves at the discontinuity points $x = -L/2$ and $x = L/2$ is purely arbitrary.
Consequently, it is important to reinforce the argument that such a possibility of interference leading to different phase time results
is strictly related to the idea of using (or not) the multiple peak (de)composition in the region where the potential barrier is localized.  
\begin{figure}[th]
\centerline{\psfig{file=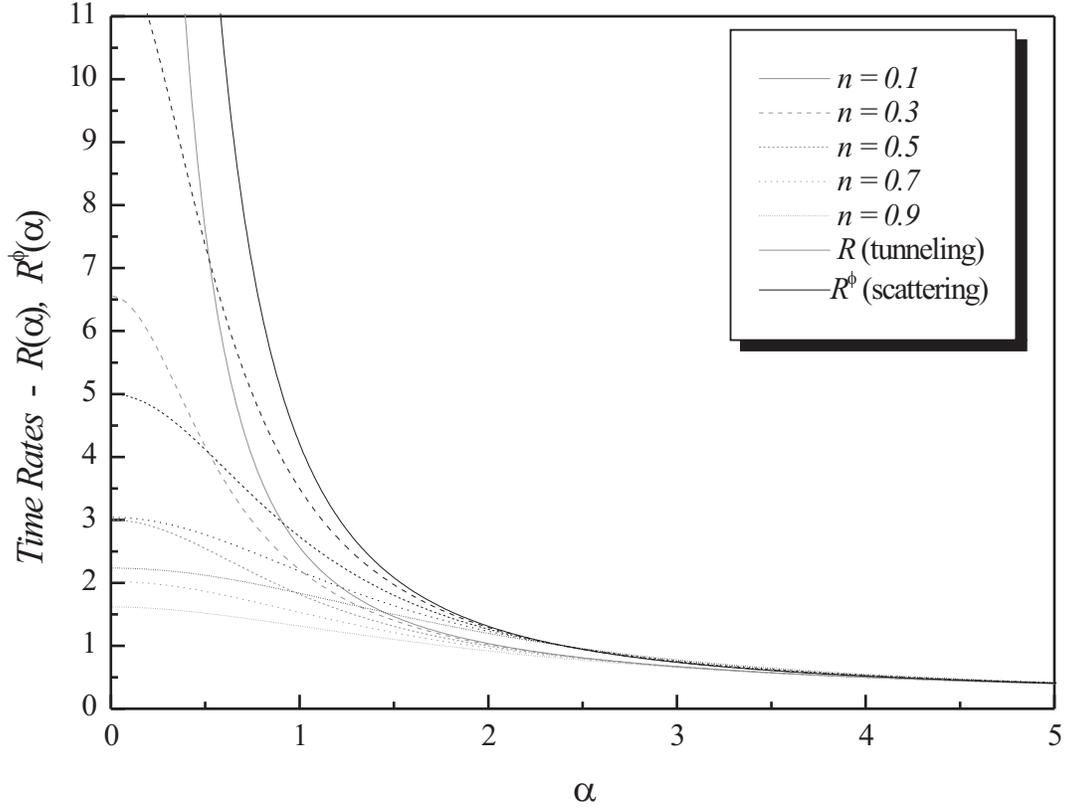,width=16cm}}
\caption{Time rates for the {\em standard} tunneling and the {\em new} scattering process.
The rates $R(\alpha)$ and $R^{\phi}(\alpha)$ can be understood as transmitted times in the units of the classical propagation time $\tau$.
Both present the same asymptotic behavior, which, in a totally restrictive mathematical sense, and in the stationary phase analysis context, offers the possibility of a {\em superluminal} interpretation for the peak of the transmitted wave packet so that the SPM can be {\em accurately} applied.
\label{fig3A}}
\end{figure}
To illustrate the difference between the standard {\em tunneling} phase time $t_{T}$ and the alternative {\em scattering} phase time $t^{^{\varphi}}_{T}$ we introduce the new parameter $n = k^{\2}_{\mbox{\tiny max}}/w^{\2}$ and we define the {\em classical} traversal time $\tau = (m L) /k_{\mbox{\tiny max}}$.
Then we can obtain the rates
\small\begin{equation}
R_{T}(\alpha) = \frac{t_{T}}{\tau}= 
\frac{2}{\alpha}\left\{
\frac{\cosh{(\alpha)}\sinh{(\alpha)} - \alpha\,n\left(2 n - 1\right)}{\left[4 n \left(1 - n\right)+\sinh^{\2}{(\alpha)}\right]}
\right\}
~~\mbox{and}~~
R^{^{\varphi}}_{T}(\alpha) = \frac{t^{^{\varphi}}_{T}}{\tau}= 
\frac{2}{\alpha}\left\{\frac{n\, \alpha + \sinh{(\alpha)}}{2n - 1 +\cosh{(\alpha)}}
\right\}\label{517}
\end{equation}\normalsize
which are plotted in the Fig.(\ref{fig3A}) for some discrete values of $n$ varying from 0 to 1.,
The most common limits of the above expressions are given by
\small\begin{equation}
\lim_{\alpha \rightarrow \infty}
{\left\{R^{^{\varphi}}_{T}(\alpha)\right\}}
= 
\lim_{\alpha \rightarrow \infty}
{\left\{R_{T}(\alpha)\right\}}
=0,~~
\label{518}
\end{equation}\normalsize
and
\small\begin{equation}
\lim_{\alpha \rightarrow 0}
{\left\{R_{T}(\alpha)\right\}}
= 1+ \frac{1}{2 n},
~~\mbox{and}~~
\lim_{\alpha \rightarrow 0}
{\left\{R^{^{\varphi}}_{T}(\alpha)\right\}}
= 1+ \frac{1}{n}
\label{519}
\end{equation}\normalsize 
Both present the same asymptotic behavior, which, in a totally restrictive mathematical sense of the stationary phase analysis context, allows the possibility of a {\em superluminal} interpretation for the peak of the transmitted wave packet.
The main point is that, by now, from the point of view of the analytical limitations, the SPM can be {\em accurately} applied.
At this point it is convenient to notice that the superluminal phenomena, observed in the experiments with tunneling photons and evanescent electromagnetic waves \cite{Nim92,Ste93,Chi98,Hay01}, generated a lot of discussions on relativistic causality.
In fact, superluminal group velocities in connection with quantum (and classical) tunnelings were predicted even on the basis of tunneling time definitions more general than the simple Wigner's phase time \cite{Wig55}.
Olkhovsky {\em et al.} discuss a simple way of understanding the problem \cite{Olk04}.
In a {\em causal} manner, it might consist in explaining the superluminal phenomena during tunneling as simply due to a {\em reshaping} of the pulse, with attenuation, as already attempted (at the classical limit) \cite{Gav84}.
The later parts of an incoming pulse are preferentially attenuated, in such a way that the outcoming peak appears shifted towards earlier times even if it is nothing but a portion of the incident pulse's forward tail \cite{Ste93,Lan89}.
In particular, we do not intend to expand on the delicate question whether superluminal group velocities can sometimes imply superluminal signaling.
It is a controversial subject which has been extensively explored in the literature about the tunneling effect (\cite{Olk04} and references therein).

Turning back to the scattering time analysis, we can observe an analogy between our results and the results interpreted from the Hartman Effect (HE) analysis \cite{Har62}. 
The HE is related to the fact that, for opaque potential barriers, the mean tunneling time does not depend on the barrier width.
For large barriers the effective tunneling velocity can become arbitrarily large so that the tunneling phase time becomes independent of the barrier width.
It seems that the penetration time, needed to cross a portion of a barrier, in the case of a very long barrier starts to increase again—after the plateau corresponding to infinite speed—proportionally to the distance\footnote{The validity of the HE was tested for all the other theoretical expressions proposed for the mean tunneling times \cite{Olk04}.}.
Our phase time dependence on the barrier width is similar to that which leads to Hartman interpretation as we can infer from Eqs.~(\ref{518}-\ref{519}).
Only when $\alpha$ tends to $0$ we have an explicit linear time dependence on $L$, 
\small\begin{equation}
t^{\varphi}_T = \frac{2\, m \, L}{w} \left(1 + \frac{1}{n}\right)
\label{14B},
\end{equation}\normalsize
which agrees with calculations based on the simple phase time analysis where $t_T = \frac{2\, m \, L}{w} \left(1 + \frac{1}{2n}\right)$, as we can observe in the by Eq.~(\ref{14}) for $n=1$.

Here at once it is important to emphasize that the wave packets for which we compute the phase times illustrated in the Fig.(\ref{fig3A}) are not constructed with the same momentum distributions.
The phase $\Theta(k, L)$ appears when we treat separately the momentum amplitudes $g(k - k_{\0})\,|T(k, L)|$ and $g(k - k_{\0})|R(k, L)|$ and the other one $\varphi(k, L)$ appears only when we sum the amplitudes $g(k - k_{\0})\,|T(k, L) + R(k, L)| = g(k - k_{\0})$ in order to obtain a symmetrical distribution. 
It {\em requalifies} the SPM for accurately computing the time dependence of the position of the peak of a wave packet. 
Moreover, some authors have correctly considered for a sufficiently complete analysis of the violations of the HE,
not only the filter action but also the spreading of the wave packets caused by the square-law dependence of the kinetic energy on $k$
(sometimes carrying into even negative tunneling times) \cite{Olk04} and, for some particular configurations of two (and more) barriers \cite{Olk02B},  
the influence of all possible resonances.
Some additional anomalies with the HE had already been discussed in the last review on tunneling time analysis \cite{Olk04}.
In spite of quoting the superluminal interpretation present in the literature since a long time ago, our discussion concerned with the definition of the strict mathematical conditions which limit the applicability of the stationary phase method for which the Hartman interpretation is valid \cite{Har62}.
Strictly speaking, the discussion of superluminal phenomena is justified only for (ultra)relativistic particles \cite{Olk92,Jak98}, for instance photons, but not for the tunneling analysis in the case of the non-relativistic Schroedinger equation.
In fact, up to now, the most interesting experiments concerning the time analysis of tunneling processes had been fulfilled with photons.

In more general lines, there have also been some trying of yielding complex time delays for tunneling analysis, ultimately due to a complex propagation constant.
In such a framework, the supposition of superluminal features is considered artificial since the transmitted peak is not causally related to the corresponding incident peak.
In certain sense it has caused some controversies with denying the physical reality to an imaginary time \cite{Lan94}.
In parallel to the most sensible candidate for tunneling times \cite{Hau89,Lan94}, a phase-space approach has been use to determine a semi-classical traversal time \cite{Xav97,Sok90,Sok94}.
This semi-classical method makes use of complex trajectories which, in its turn, enables the definition of real traversal times in the complexified phase space \cite{Xav97,Sok90}. 
It is also commonly quoted in the context of testing different theories for temporal quantities such as arrival, dwell and delay times \cite{Hau89,Lan94} and the asymptotic behavior at long times \cite{Jak98,Bau01}.
In particular, it suggests that the idea of complexifying time should be investigated for some other scattering configurations, which reinforces the more general assertion that the investigation of wave propagation across a tunnel barrier has always 
been an intriguing subject which is wide open both from a theoretical and an experimental point of view.

As a possible solution for partially overcoming some of the incongruities here pointed out and quantified, which appear when we compute tunneling phase times in the SPM framework, we have claimed the relevance of the use of the multiple peak decomposition \cite{Ber04} technique previously developed for the above barrier diffusion problem.
Essentially, we have introduced a way for comprehending the conservation of probabilities for a very particular tunneling configuration where the asymmetry presented in the standard case was eliminated, and the phase time could be accurately calculated.    
We mention for a subsequent analysis the suggestive possibility of investigating the validity of our approach when confronted with the intriguing case of multiple opaque barriers \cite{Esp03}, in particular, in the case of non-resonant tunneling.
Still concerned with subsequent theoretical perspectives, the symmetrical colliding configuration also offers the possibility of exploring some problems involving soliton structures.
To conclude, all the above arguments reinforce the necessity of searching the appropriate framework where barrier traversal times can be computed in the most generalized way.

{\bf Acknowledgments}
We would like to thank FAPESP (PD 04/13770-0) for the financial support.
We also thank the referee of this manuscript for his/her useful and very constructive comments
which certainly contributed to the improvement of the text.

\end{document}